# Adversity Index for Clinical Trials

## An Inclusive Approach for Analysis of Safety Data

Sharayu Paranjpe[1], Anil Gore[1]

1. Cytel Statistical Software and Services, Pvt. Ltd. Pune, India

Correspondence to: Sharayu Paranjpe

sharayu.paranjpe@cytel.com,

Tel: +91 20 6709 0189

Address: Cytel Statistical Software and Services, Pvt. Ltd.

Lohia-Jain IT Park, Paud Road, Pune 411 038

*'... seeing an analogy where no one saw one before'*
*(Arthur Koestler, The Act of Creation, p. 175)*

## Abstract:

Safety and efficacy are two major aspects in assessing worth of a new health intervention. Statisticians tend to emphasize inference on efficacy. In comparison, inference on safety receives less attention. This is mainly because of complexity of safety data. This complexity level is comparable to that in biodiversity data in ecology. We draw analogy from biodiversity literature to summarize adverse event (AE) data. This article proposes an index named 'adversity index' (AdX for short), which combines data on all adverse events encountered in a clinical trial and not just the ones of 'special interest'. In this sense the approach is inclusive. In ecology this index is known as 'Shannon- Wiener' diversity index. AdX follows asymptotic normal distribution which permits use of standard statistical tests for treatment comparisons. AdX is a simple measure of risk quantification, which is lacking in current literature. This single number summary is shown to facilitate safety profile comparison of treatments, overall as well as by subgroups. Further, it is shown how the index can be used for decision making by interim review committees like DSMB/DMC who find it difficult to take a go/ no-go decision in a short time span. Regulators can use AdX as a quantitative measure of risk while assessing benefit risk balance. This approach can sometimes lead to results that would have been missed without it. These ideas are illustrated using anonymized data on Phase III clinical trials conducted by major pharmaceutical companies; two on breast cancer and two on diabetes.

## 1. Introduction:

Safety and efficacy are two major parameters in assessing the worth of a new health intervention. *'Drug approval decisions are based on a comprehensive assessment of the benefits of the drug and its known and potential risks'* [1]. Statisticians tend to focus more on inference regarding efficacy than safety. Current practice of inference on safety is mainly based on adverse events of special interest, generally a tiny fraction of total AE types recorded while remaining bulk of adverse events are reported only as counts and percentages. This article proposes an index named 'adversity index' (AdX), which represents integration of data on all adverse events encountered in a clinical trial. Here we demonstrate that this single value summary facilitates the safety profile comparison of treatments, at an overall and subgroup level. Further, we illustrate the potential use of this index by the Data Safety Monitoring Board at the time of interim review. In addition, the index can serve as a simple measure of overall risk, which is lacking in the current literature and can be used by the regulators while assessing benefit risk balance. AdX based analysis can sometimes reveal results that would not have been noticed with conventional approach. The proposed approach uses AE data more intensively than current methods; however, it is not meant to replace but supplement the conventional method of focusing on events of special interest. If a treatment is judged to be unsafe and unacceptable based on rates of occurrence of these events (as in case of Thalidomide or Vioxx) other adverse events may not play a role in decision making. Proposed approach will play a useful role in all clinical trials excluding such cases.

## 2. Outline of the paper:

Section three provides motivation of the paper. It describes current practice of safety data reporting in clinical trials and points out its limitations. It draws analogy between complexity of adverse event (AE) data in clinical trials and biodiversity data in ecology. Using this analogy, an index of biodiversity is proposed as a summary measure of AE data. Interpretation of and statistical inference on this summary measure are explained. Various uses of this index are illustrated in section four through analysis of Phase III clinical trials conducted by pharmaceutical companies.

## 3. Safety Data Analysis: Current scenario and proposed modification

### 3.1 Goals of safety data analysis:

In a typical clinical trial, an individual adverse event (AE) is noted immediately and acted upon as necessary. Collectively, safety data are examined at multiple stages: (a) by the DSMB/DMC during interim review of the trial to judge whether safety concerns are serious enough to warrant stopping the trial, (b) by the sponsors at the end of the trial to prepare clinical study report (CSR) following ICH E3 guidelines, and (c) by the regulators for benefit risk assessment prior to decision on drug approval.

### 3.2 Limitations of current practice:

In stage (a), a statistician providing data to a DSMB/DMC faces a dilemma: reporting all the data from the case report forms may overwhelm the committee members; on the other hand, selective reporting may hamper the ability of the committee to make informed judgments about safety. Generally the committee is presented with bulky and undigested data. The task of interim review by the DSMB/ DMC is complicated by the extremely limited time available to *'wade through seemingly endless pages'*[2]. This situation warrants correction. The proposed approach offers one solution to this problem.

In stage (b), the safety data analysis in a typical CSR is based on well-established statistical methods. The trial protocol often specifies certain adverse events (AE) being of 'special interest' in relation to disease condition and drug under study, called Tier 1 AEs[3]. Rates of occurrence of these events are compared across treatments using chi-square test, Fisher's exact test etc. Sometimes in literature we come across analysis of Tier 2 events as well, similar to analysis of Tier 1 events. In either case the analysis is beset with the problem of multiple testing and false discovery rate (FDR)[4]. Occasionally one encounters suggestions to use graphical tools to summarize safety data[5,6]. They do not seem to have a wide acceptance. In general, there is awareness that '*Statistical methods for drug safety assessment are still evolving*'[7]. Current methods fail to provide a cogent summary of the safety data. A typical CSR offers a 'safety data summary' which runs into hundreds of tables, with each table possibly spanning across multiple pages. In a breast cancer trial by Eli Lilly (NCT00006459), the length of section 14.3, 'Safety Data Summary' of the CSR was more than 220 pages. Such a length contradicts the concept of summary. In case of the diabetes trial by Boehringer Ingelheim (NCT01159600) the safety tables span across more than 1800 pages. The absence of a clear and concise summary of safety data poses a challenge in weighing it against benefit. In place of this bulk what is really needed is a concise measure of overall safety (profile) of a treatment.

At stage (C) regulators involved in benefit risk assessment also face a problem. Review of current benefit-risk literature reveals a common difficulty in quantifying overall risk. As a result *'There is no defined and agreed methodology to combine benefits and risks to allow direct comparisons.'*[8].

In addition to the above three stages, a special situation requiring focussed safety data analysis may arise. Here is an illustration of such a situation: *'Either before or after marketing approval, there is sometimes a need for a large randomized controlled safety study to evaluate a concern that may have arisen from observational adverse event reporting'*[9]. The methods described below will also be applicable to such a situation.

### 3.3 Root cause of weakness in current practice:

Statistical methods for demonstrating efficacy of a new medicinal product are well developed. Why do methods for safety data analysis lag behind those for efficacy evaluation? In efficacy evaluation, the assessment is relatively straightforward because while developing a drug or a treatment, the researcher knows precisely the type of benefit that the treatment should generate. For example, in case of an antihypertensive drug lowering of blood pressure is desired. Similarly, for anti diabetic drug lowering of blood sugar or for an oncology study, increase in median overall survival is desired. In addition to these primary end points, a relatively small set of co-primary or secondary efficacy end points may also be of interest. Thus the number of endpoints to be assessed for efficacy is limited, and the corresponding statistical methods for treatment comparison are well established.

In contrast, when we consider the safety aspect of the drug in general and adverse events in particular, the complexity level is high: the number of distinct AEs (AE types) in a trial is large (several hundred), the severity/ seriousness levels of AEs are different, the same subject may experience multiple AEs, the same AE may affect many subjects, the same AE may affect the same subject repeatedly, AEs may be associated etc. Therefore, statisticians struggle to offer a meaningful summary. This could create a feeling of despondency when it comes to safety comparison of treatments, which is reflected in the following comment : *'The FDA, industry and academia remain in a quandary as to how to respond in a responsible fashion to observed differences in reported frequencies of adverse events.'*[10].

Sometimes, complications may arise in the analysis of efficacy data as well. The recent ICH E9 (R1) addendum discusses the difficulty of accommodating inter-current events. It suggests strategies for developing estimands to handle such events. The current debate on estimands mainly revolves around efficacy evaluation. However, the concept is also applicable to safety endpoints and different safety estimands may be of interest. We see an opportunity for developing a new approach to summarising and reporting AE data using an estimand.

### 3.4 Proposed solution based on analogy between safety data and biodiversity data:

Since conventional methods do not seem to give satisfactory results, an unconventional line of attack may be productive. A similar situation may have been encountered in a completely different branch of science and a solution may have been found there. In such a case, drawing an analogy and borrowing relevant tools may be helpful. This is exactly what we propose to do.

We draw an analogy between AE data and biodiversity data. The complexity of adverse event (AE) data in a typical clinical trial is comparable to the complexity of biodiversity data. In ecology, the measurement of biodiversity is a major topic of interest. Biodiversity broadly means variability in life forms. In a typical ecosystem, e.g. a tropical forest, many types of organisms coexist. These organisms include trees, insects, mammals, birds etc. They vary in numbers, types and other parameters. Ecologists are interested in quantifying this variability. Let us consider only birds. Ecologists study birds in a forest through a sample survey of the field by recording the bird species encountered and their count (abundance). This information forms the basic data set. In a tropical forest, the list of bird species can be very long (in hundreds) and the counts can range from one individual to several thousand individuals. The tool used by ecologists to summarise this variation is the so called 'index of diversity'. A forest with a higher index of diversity is considered to be richer. Diversity can be assessed at different taxonomic levels (species, genus, family etc.).

The basic data for the measurement of species diversity is a list of species along with the corresponding abundances. The parallel for clinical trials is a list of AE types along with the corresponding frequencies of occurrence in a trial. These AE data represent the safety profile of a treatment. This parallel suggested a possible deeper analogy. Table 1 specifies the other components in the analogy between AE data and biodiversity data.

| Table 1: Analogy between biodiversity data and AE data ||
|---|---|
| **Biodiversity Data** | **Adverse Event Data** |
| Field Survey | Clinical trial or a set of clinical trials (ISS) |
| Species | AE type |
| Species Abundance | Frequency of occurrence of an AE type |
| Group of Species | AEs by SOC/ any appropriate group |
| Sampling effort | # Patients / patient days in a trial |
| Species turnover (α, β, γ diversity) | AE turnover across treatments (intersection sets) |
| Taxonomic hierarchy | MedDRA hierarchy |

This remarkable similarity between the two domains has prompted us to explore the possible use of summary measures from ecology for AE data in clinical trials. Ecologists use not one but many diversity indices. All these indices are based on the number of species (say K) and their relative abundances ($p_i$, i= 1 to K, $\sum p_i =1$). Two of these indices are well known in ecology literature: Shannon- Wiener index (SW) and Simpson's index (SI). The SW index is our choice. This index originated in thermodynamics and was later adopted in other branches of science such as chemistry, biology, linguistics etc. often under the name 'Entropy'. A relevant example is its very recent use in the anonymization of clinical trial reports[11]. Here uncertainty of re-identification of 'protected personal information' (PPI) is important. SW index captures this uncertainty. Greater the value of index, greater is the uncertainty. The formula for the Shannon-Wiener index is

$$SW = -\sum p_i * ln(p_i) \qquad \text{Eq.1}$$

where $p_i$ is the relative abundance of the $i^{th}$ species and the sum is over all species. Higher value of the index indicating greater randomness or greater 'disorder' represents a 'richer' ecosystem. The value of index depends on two factors: first, larger number of species leads to larger value of index; second, for a given number of species, evenness in abundance across species results in a higher value of the index. To understand the effect of evenness on the index, consider a hypothetical case of three communities for comparison; each of the three communities, has five species (K= 5) with a total of 100 individuals (N=100). The abundances of individual species and the resulting diversity indices are shown in Table 2.

| Table 2: Species-wise abundances in three hypothetical communities and their diversity indices ||||||||
|---|---|---|---|---|---|---|---|
| Community | Species ||||| Index SW | Comment |
| | S1 | S2 | S3 | S4 | S5 | | |
| 1 | 1 | 1 | 1 | 1 | 96 | 0.22 | Extreme unevenness |
| 2 | 1 | 3 | 6 | 10 | 80 | 0.73 | Intermediate evenness |
| 3 | 20 | 20 | 20 | 20 | 20 | 1.61 | Extreme evenness |

In the first community four species are represented by only one individual each, and the fifth species has 96 individuals. The second community has more than one individual in some species. In the third community each of the five species is represented by 20 individuals. The diversity index has the smallest value for the first community, the largest value for the third community, and an intermediate value for the second community.

### 3.5 Definition and interpretation of adversity index:

The central proposal of this paper is that AE data in a clinical trial should be summarised using the Shannon Wiener (SW) index defined in Eq. 1. Let us suppose that there are K AE types, with a total number of episodes N and the relative frequency of occurrence of $i^{th}$ AE type denoted by $p_i$. These data can be summarised using Eq. 1 and we name the resulting value as 'Adversity Index', AdX for short. As stated in section 3.4, the AdX value will increase as K increases. Further, the AdX value will increase if $p_i$ values are similar across AE types. It may be useful to emphasize that the index is based on counts of AE episodes and not on the number of subjects affected. Further it should be kept in mind that, like any summary measure or index, AdX suppresses many details of the data. Equality of AdX for two data sets does not imply identical profiles. However, this summarization facilitates comparison of treatments. Use of AdX is intended to draw attention to a treatment with more AE types and their relative abundance. It is not intended to identify individual types of AE's or rare AE's.

Conventional approach to safety data analysis makes a distinction between levels of seriousness, grades of severity, different body systems etc. So far we have not incorporated these distinctions in AdX. However, when interest is focussed on a subgroup (by age/ gender/ SOC/ seriousness/ severity etc.) index can be calculated for that subgroup. Once a subgroup shows significant treatment difference in AdX, then additional exploratory analyses and/or more careful perusal of the types of AEs are likely needed to explain the AdX difference. A common terminology in literature is Tier 1/Tier 2-3 events. This is yet another way of sub grouping. Current practice is to compare rates of individual Tier 1 AEs across treatments. Tier 2-3 events are reported as counts and percentages. AdX analysis can also be carried out for one or more of these subgroups as needed.

In the ecological context, a higher value of SW is desirable. Is this true for AdX also? We contend that the opposite is true. If there is only one AE type, value of AdX is zero. This zero value does not indicate absence of risk. It simply indicates that we know precisely where to focus the risk mitigation efforts. As the possible number of AE types increases and the AdX value increases, the challenge in preparing for all eventualities becomes more daunting. Secondly, if the number of AE types is the same for two treatments, a higher AdX implies a greater evenness in counts of AE occurrence across AE types. This scenario indicates greater uncertainty about which AE type will affect the subjects. As a consequence the risk mitigation effort will need to pay attention to many more AE types, which is an undesirable situation. Hence we regard higher value of the index as an indication of lower safety level.

**3.6 Inference on AdX and safety comparison of treatments:**

A key question of interest for sponsors or regulators is which of the treatments is better in overall safety. The simplest answer is that the treatment with a lower value of AdX is safer. The next essential question is whether the difference in AdX is statistically significant or due to chance alone. To answer this question, we have to consider the distributional properties of AdX such as standard error and confidence interval. Here, it is perhaps relevant to make a distinction between a true and unknown population AdX (PAdX) and the estimate from sample data. Consider a target population of patients with a specific disease condition (say breast cancer) treated with a specific drug (say Gemcitabine by Eli Lilly). Collection of possible adverse events in cancer patients getting this treatment is the universe of interest. The adversity index based on this set and associated probability vector is PAdX (an estimand for safety). In a particular clinical trial we get a sample of these adverse events and we calculate a sample AdX. This is the estimate of PAdX. For convenience, we will drop the prefix P from now on. The context will make it clear whether we are discussing sample AdX or unknown population AdX. If the sampling distribution of AdX is bell-shaped, statistical methods based on normal distribution can be used. AdX follows an asymptotic normal distribution[12] with mean PAdX and variance

$$\frac{\sigma^2}{N} = \frac{1}{N}[\sum_{i=1}^{K} p_i * (ln(p_i) + PAdX)^2] \qquad \text{Eq. 2}$$

where N is the total number of AE episodes, K is the number of AE types and $p_i$ is the proportion of episodes of AE type i. While using this result for treatment comparison etc. we will replace PAdX by sample quantity AdX. The standard error of the difference in AdX of two treatments T1 and T2 is

$$SE(diff) = [Var(AdX(T1)) + Var(AdX(T2))]^{\frac{1}{2}} = \left[\frac{\sigma_1^2}{N_1} + \frac{\sigma_2^2}{N_2}\right]^{\frac{1}{2}}.$$

## 4. Application of Adversity Index to data in clinical trials:

### 4.1 Summary and AdX of AE data:

The remainder of this paper illustrates the use of AdX for comparing the safety profiles of treatments in four clinical trials (coded in text as EL, BI, RO and GSK). Table 3 gives the details of these trials.

| | | Table 3: Data Source (Anonymized) | | |
|---|---|---|---|---|
| Serial Number | NCT | Sponsor | Indication | Code used in text |
| 1 | NCT00006459 | Eli Lilly | Breast cancer | EL |
| 2 | NCT01159600 | Boehringer Ingelheim | Diabetes Mellitus Type II | BI |
| 3 | NCT00333775 | Roche | Breast cancer | RO |
| 4 | NCT01128894 | GlaxoSmithKline | Diabetes Mellitus Type II | GSK |
| All 4 trials are Phase III. Data were made available by https://www.clinicalstudydatarequest.com | | | | |

As discussed in section 3.2 the summary provided in a typical CSR is not concise and is often difficult to comprehend or interpret. Our attempt is to offer a better alternative. Table 4 shows a high level summary of the various counts related to safety for all four clinical trials analysed.

| Table 4: High-level summary of four trials | | | | | |
|---|---|---|---|---|---|
| | Study | | | | |
| | EL | BI@ | | RO | GSK |
| Therapeutic Area: | Breast Cancer | DM II (Met) | DM II (Met + SU) | Breast Cancer | DM II |
| # Subjects | 521 | 718 | 786 | 736 | 812 |
| # AE episodes | 55,803 | 1052 | 1655 | 17448 | 3416 |
| # AE types | 208 | 325 | 384 | 862 | 549 |
| # Subjects with at least one AE (%) | 521 (100.0) | 426(59.3) | 541(68.8) | 728(98.9) | 667(82.1) |
| # Treatment arms | 2 | 4 | 4 | 3 | 2 |
| Test drug | Gemcitabine | Empagliflozin | | Bevacizumab | Albiglutide |
| @ The BI trial had two sub-studies based on the background therapy (i) Metformin (Met) and (ii) Metformin + sulphonylurea (Met+SU) | | | | | |

We note in passing that (i) the number of AE episodes in oncology trials is far greater than in diabetes trials, (ii) within oncology, the count of AE types in RO trial is over four times the count in EL trial even though the count of AE episodes is less than a third.

Table 5 shows our first attempt at creating an overview of the AE data for the EL study. Here we observe that all the subjects in the EL trial experienced at least one AE. Total number of AE episodes in the GT group is 21% greater than that in the T group. The number of distinct AEs is slightly greater in the GT group than in the T group.

| Table 5: Summary of AE data by treatment (EL) | | | |
|---|---|---|---|
| Variable | Treatment† | | Total |
| | GT | T | |
| # subjects | 262 | 259 | 521 |
| # AE episodes | 30446 | 25357 | 55803 |
| # distinct AEs | 187 | 178 | 208 |
| # subjects with at least one AE(%) | 262(100%) | 259(100%) | 521(100%) |
| AdX | 3.64 | 3.48 | |
| SE(AdX) | 0.0079 | 0.0086 | |
| Difference (AdXGT-AdXT) and SE | 0.16(0.0117) | | |
| Data Source: NCT00006459; Sponsor Name: Eli Lilly; Indication: Breast Cancer. †GT= Gemcitabine + Paclitaxel; T= Paclitaxel. | | | |

While the above summary appears concise and relevant, many clinical trials have features which require a more detailed table. The EL trial on breast cancer naturally involves only one gender. In general, there may be interest in gender-wise comparison of response to treatments. Other features of interest can be age, background therapy etc. An appropriately modified version of the above table can be used.

Turning to inference, the Z-statistic for comparison of AdX for two treatments in the EL trial is 0.16/0.0117= 13.67, which is statistically significant at every reasonable choice of (α); level of significance. We note that standard errors of AdX for both treatments and their difference are very small because of large values of N1 (30446) and N2 (25357). Thus, with large counts of AE episodes, the sample AdX values are essentially the population values. Such large number of AE episodes may not be a common occurrence. To see results for a case with moderate number of AE episodes we consider the BI trial. It had four treatment arms compared separately in two sub-studies (by background therapy). Table 6A shows AdX values for each gender, sub study and treatment. Table 6B gives p-values for treatment comparison within each gender and sub study.

| Table 6A: AdX and SE by treatment and gender (BI) | | | | | |
|---|---|---|---|---|---|
| **Background Therapy** | **Gender** | Treatment arm | | | |
| | | 10 mg | 25 mg | Placebo (P) | Open 25 mg |
| Metformin | Female | 4.38 (0.0654) | 4.25 (0.0657) | 3.97 (0.0793) | 3.02 (0.0921) |
| | Male | 4.07 (0.0859) | 4.25 (0.0704) | 4.32 (0.0696) | 3.63 (0.0591) |
| Metformin + SU | Female | 4.04 (0.0836) | 4.01 (0.0912) | 4.05 (0.0732) | 3.66 (0.0861) |
| | Male | 4.08 (0.0993) | 3.69 (0.1116) | 4.16 (0.0806) | 3.84 (0.0836) |
| Data Source: NCT01159600; Sponsor Name: Boehringer Ingelheim (BI); Indication: Diabetes Mellitus Type II. | | | | | |

| Table 6B: p-values for Comparison of treatments within each sub study (BI) | | | | | |
|---|---|---|---|---|---|
| Sub Study | Gender | 10 mg - P | 25 mg -P | 25mg (Open) –P | (Blinded –open) 25 mg |
| Met | Females | <0.001 | <0.003 | <0.001 | <0.001 |
| | Males | 0.012 | 0.240 | <0.001 | <0.001 |
| Met + SU | Females | 0.46 | 0.36 | <0.001 | 0.003 |
| | Males | 0.266 | <0.001 | 0.003 | 0.141 |
| Data Source: NCT01159600; Sponsor Name: Boehringer Ingelheim (BI); Indication: Diabetes Mellitus Type II. | | | | | |

Notice that 5 out of the 16 comparisons fail to attain significance. Remaining 11 comparisons yield low p-values. It is generally expected that higher dose is accompanied by higher toxicity. So AdX for 25 mg should be higher than that for placebo. In fact in case of males with metformin as background therapy, AdX for 25 mg is comparable to placebo. Even more surprising is the case of males with Met + SU as background therapy where AdX for 25mg dose is significantly smaller than placebo, indicating better safety. The corresponding picture in females is exactly the opposite. Such subtle features seem to be missed out in conventional analysis in CSR. Here is what CSR reports: *'Overall empagliflozin treatment was generally well tolerated and showed similar safety profile compared to placebo in patients with Metformin only and Metformin plus sulphonylurea background medications (CSR p. 19)'.*

## 4.2 Clinical significance of AdX differences:

In clinical trials it is prudent to check the clinical significance along with the statistical significance. How can the clinical significance of difference in AdX be judged? AdX is only an index; therefore its absolute value is hard to interpret. Ecologists have faced the same difficulty. Hence it is relevant to examine their approach. Ecologists use a concept called 'effective number of species'. Analogously we will introduce the concept of **E**ffective **A**dversity **L**oad **S**core (EALS). EALS is a transformation of AdX.

We explain the concept of EALS with an example (Table 7). Consider two treatment arms. In the first arm, there are four AE types (K=4), with the corresponding frequencies of occurrence shown in column $N_{1i}$. In the second arm, there are two AE types (K* =2). In both cases, the AdX value is the same, viz.0.69. Therefore, the two safety profiles are similar (in terms of AdX). The main difference is that in Arm 2 both AEs occur with equal frequency. The second arm is synthetic and is not expected to be encountered in reality. However, there is a mathematical relationship between K* and AdX. The value 2 is equal to exp (0.69). Therefore, the Effective Adversity Load Score (EALS) for Arm one is 2.

| Table 7: Example data with EALS =2 | | | | |
|---|---|---|---|---|
| | Arm 1 | | Arm2 | |
| AE type | $N_{1i}$ | $P_{1i}$ | $N_{2i}$ | $P_{2i}$ |
| AE1 | 81 | 0.81 | 50 | 0.50 |
| AE2 | 7 | 0.07 | 50 | 0.50 |
| AE3 | 6 | 0.06 | 0 | 0.00 |
| AE4 | 6 | 0.06 | 0 | 0.00 |
| Total | 100 | 1.00 | 100 | 1.00 |
| AdX | | 0.69 | | 0.69 |

In general the formula K* = exp (AdX) may not yield an integer value. Thus, it need not represent any real treatment. K* is a mathematical characterization of AdX. If K* happens to be an integer, we can imagine a hypothetical treatment with K* AE types each occurring with the same frequency. To phrase it differently, a treatment with K AE types with unequal frequencies of occurrence is in effect equivalent to another (hypothetical) treatment with K* AE types each with the same frequency of occurrence. The safety profile of a treatment can be represented either by AdX, or, equivalently, by K*, which we call EALS. We believe that the difference in EALS value may be easier for clinical interpretation.

For the EL trial, let us use the AdX values in Table 5 to calculate the EALS. Here, EALS (T) =32 and EALS (GT) = 38 (values rounded). The difference [AdX (GT) – AdX (T)] = 3.64-3.48 = 0.16 translates into increment of six equally frequent AE types. For the BI trial (Table 6A) consider the 25 mg and placebo groups for males of the Met + SU sub-study. The EALS for placebo is [exp( 4.16)] =64 and that for 25 mg is [(exp(3.69)]= 40. Thus, the safety profile of 25 mg group is as if subjects experienced 24 fewer (effective) number of AE types compared to subjects in placebo group. Perhaps clinicians can relate more easily to this type of statement.

### 4.3 Normalisation of AdX:

The above approach is suitable for comparing treatments when the number of AE types recorded in two treatments is similar. In some special scenarios (as illustrated below) when this is not so, a further normalization becomes necessary. In such cases, if AdX values are different, a comparison of EALS values is not sufficient. Hence we divide each EALS value by the corresponding value of K, observed number of AE types. We call this value Standardised EALS or SEALS (=K*/K). Note that K* is always less than K; therefore, the value of SEALS will always be between 0 and 1. A value of 1 indicates K* = K, i.e. all AE types occur with roughly equal frequency in data. This scenario poses a challenge for safety risk management. In general, the smaller the value of SEALS, the easier is the safety risk management.

Here is an example where the values of K are very different in two groups. In the BI study on Type II diabetes the same dose of 25 mg of empagliflozin is administered as a blinded arm to one group and as an open arm to another group.

This feature of the BI trial can be considered as an experiment by itself. Blinding is an essential aspect of a clinical trial design. Regulators not only insist on blinding, but also demand an assessment of the success of blinding. This requirement is because of a clear recognition that successful blinding reduces the placebo effect and the possible bias in efficacy assessment. What is the impact of blinding on safety assessment? It is hard to find in literature any discussion about the relation between blinding and safety assessment. Further we do not often encounter the same treatment in a clinical trial under both blinded and open-label conditions. What we do encounter is an 'open-label extension' that follows a blinded trial but here *'Analysis strategies need to be developed and implemented to provide unbiased estimates of safety and tolerability'*[13].

By design in the BI trial, the open-label arm has considerably fewer subjects (about a third of blinded arm). When the number of subjects is smaller, rare events are easily missed. Hence the number of observed AE types is smaller for open arm. Table 8 shows the dilemma encountered and a solution.

| Table 8: AdX normalization by number of AE types (K) (BI Study) | | | | | | | | |
|---|---|---|---|---|---|---|---|---|
| | AdX(25 mg) | | EALS (K*: rounded) | | # AE types (K) | | SEALS (K*/K) | |
| 1 | 2 | 3 | 4 | 5 | 6 | 7 | 8 | 9 |
| | Blinded | Open | Blinded | Open | Blinded | Open | Blinded | Open |
| MetF | 4.25 | 3.02 | 70 | 20 | 95 | 23 | 0.74 | 0.89 |
| MetM | 4.25 | 3.63 | 70 | 38 | 92 | 41 | 0.76 | 0.92 |
| MSF | 4.01 | 3.66 | 55 | 39 | 107 | 48 | 0.51 | 0.81 |
| MSM | 3.69 | 3.84 | 40 | 47 | 105 | 60 | 0.38 | 0.78 |
| Data Source: NCT01159600; Sponsor Name: Boehringer Ingelheim (BI); Indication: Diabetes Mellitus Type II. | | | | | | | | |

This trial consists of two sub-studies depending on the background medication (Metformin and Metformin + sulphonylurea). We get four groups because there are two background medications (Met/ MS) and two genders (M/F). Columns 2 and 3 of Table 8 show that in three out of four cases the blinded arm has a higher AdX suggesting lower safety level. We transform these values to EALS values shown in columns 4 and 5. Here for the Metformin female group, the blinded arm has (70-20=) 50 additional effective AE types compared to the open-arm. This difference appears to be large. Columns 6 and 7 point out that the number of AE types in the blinded arm is much greater larger than the corresponding number in the open-arm. Therefore, further normalization is necessary. It is shown in columns 8 and 9.

Remarkably, after normalization, the picture reverses. In all four groups, the SEALS value for the open arm is greater than that for corresponding value in blinded arm. It suggests that open arm is less safe than blinded arm. In conclusion the ranking based on AdX or EALS should be used only when the number of AE types is similar. Otherwise use of SEALS is recommended.

### 4.4 Subgroup analysis:

Protocols often provide for analysis of efficacy endpoints in subgroups by demographic factors, prognostic factors, prior treatment etc. Use of AdX allows us to carry out similar analysis for safety. An illustration of AdX by age groups for the RO trial is given in Table 9 which shows that the youngest and the oldest age groups respond differently than the intermediate age groups. In both the extreme age groups, the high dose has a significantly higher AdX value than the low dose or placebo.

Perhaps there is awareness about safety issues in case of the age group '> 65'. It is reflected in various comments in the CSR. For example *'Some side effects are more common in elderly patients than in younger patients'.* (p. 874). In addition to this general reference, specific adverse events are also mentioned. For example *'overall increase in bleeding events'* (CSR p 737), *'increased risk of ...CVAs/TIAs/ MIs'* (CSR p 379).

| Table 9: AdX by age group and treatment (RO) | | | | | |
|---|---|---|---|---|---|
| | | Age group (Years) | | | |
| | Treatment | <40 | 40-50 | 50-65 | >65 |
| AdX | Doc | 4.19 | 4.58 | 4.67 | 4.5 |
| | 7.5Bv | 4.21 | 4.61 | 4.68 | 4.4 |
| | 15Bv | 4.45 | 4.55 | 4.72 | 4.65 |
| p-value (Diff) | 7.5Bv-Doc | > 0.05 | > 0.05 | > 0.05 | > 0.05 |
| | 15Bv-Doc | <0.001 | > 0.05 | > 0.05 | <0.001 |
| | 15Bv-7.5Bv | <0.001 | > 0.05 | > 0.05 | <0.001 |
| Data Source: NCT00333775; Sponsor Name: Roche; Indication: Breast cancer. | | | | | |

However, the suggestion in Table 9, that a high dose is less safe for the age group below 40 years appears to be new. It is brought out by use of AdX.

This is not a unique instance. Next example shows subgroup difference in overall safety when CSR suggests otherwise. The CSR of the BI trial gives the following comment on safety analysis: '*Overall empagliflozin treatment was generally well tolerated and showed similar safety profile compared to placebo in patients with Metformin only and Metformin plus sulphonylurea background medications*' (CSR p. 19). In contrast Table 6A suggests that there may be treatment differences. Consider the female group with Metformin as background medication. Here the AdX value for placebo (3.97) is less than either of active treatment arms (4.38 and 4.25). Further the differences are statistically significant (first row of Table 6B). Why was this difference missed out in CSR? Perhaps the explanation lies in the second row of Table 6A (for males). Here the AdX value for placebo (4.32) is greater than either of active treatment arms (4.07 and 4.25). Further the difference between 10 mg and placebo is significant and active treatment is safer than placebo. This is a reversal from the female group. Perhaps because CSR gives results for data pooled over genders, the opposing differences get averaged out.

### 4.5 AdX by SOC:

An AE episode is routinely classified by System Organ Class (SOC). Depending on the therapeutic area one or more specific SOCs may be of interest. For example the CSR of the EL trial on oncology states: '*As expected with chemotherapy, clinically significant laboratory toxicities were primarily hematologic*' (EL CSR p. 5). Other therapeutic areas may have some other SOCs of primary interest. Therefore the safety comparison of treatments by SOC is useful. Hence SOC wise AdX can potentially reveal more insights. We use the BI trial data to illustrate this possibility. Table 10A shows treatment comparison of AdX for each SOC separately. Last two columns give the p-values for comparison of active treatment with placebo. The differences are significant in two cases; 'Infection and infestations' (10 mg Vs placebo) and 'Metabolism and Nutrition' (25 mg Vs Placebo).

| Table 10A: SOC wise comparison of AdX (BI-Metformin group) | | | | | |
|---|---|---|---|---|---|
| | Treatment (AdX and SE) | | | p-value (diff) | |
| SOC | 10 mg | 25 mg | Placebo | 10mg-Placebo | 25mg-Placebo |
| GI | 2.43(0.127) | 2.37(0.157) | 2.3(0.155) | 0.258 | 0.375 |
| I & I | 2.82(0.126) | 2.44(0.130) | 2.5(0.135) | 0.042 | 0.626 |
| MAN | 1.84(0.169) | 2.11(0.132) | 1.49(0.176) | 0.076 | **0.002** |
| MS | 2.32(0.104) | 2.36(0.133) | 2.44(0.136) | 0.759 | 0.663 |
| CNS | 2.04(0.199) | 2.02(0.171) | 1.71(0.246) | 0.148 | 0.150 |
| Renal | 1.68(0.163) | 2.03(0.107) | 1.89(0.201) | 0.792 | 0.270 |
| Data Source: NCT01159600; Sponsor Name: Boehringer Ingelheim (BI); Indication: Diabetes Mellitus Type II. | | | | | |

| Table 10B: # AE episodes in MAN by treatment (BI-Metformin group) | | | |
|---|---|---|---|
| AE(total # types 15) | 10 mg | 25 mg | Placebo |
| Hyperglycaemia | 6 | 2 | 23 |
| Hypoglycaemia | 21 | 1 | 1 |
| Others | 18 | 19 | 17 |
| AE with zero count | 3 | 5 | 6 |
| Total | 45 | 22 | 41 |
| Data Source: NCT01159600; Sponsor Name: Boehringer Ingelheim (BI); Indication: Diabetes Mellitus Type II. | | | |

Next step is to look for individual AE types responsible for significant differences. Let us select SOC MAN (metabolism and nutrition), the case with lowest p-value. The results are in Table 10B. There were 15 distinct AE types in this SOC. Out of these, 2 AE types have substantial counts in at least one treatment arm. Hyperglycaemia has the highest count in the placebo group while hypoglycaemia has the highest count in the 10 mg group. Identifying such specific events is precisely what a DSMB looks for. In this sense, SOC wise AdX analysis can be of assistance to DSMB.

Again this is not a unique case. In the RO trial SOC wise comparison of treatments yields only one significant difference. It is for the SOC 'Respiratory, thoracic and mediastinal disorders'. Drilling down to individual AE types reveals that only one AE type stands out viz. Epistaxis. Here active treatment arms have much larger episode counts than placebo.

### 4.6 Safety picture at interim looks:

Above illustration is based on data at the end of the trial; but similar analysis is possible at interim looks as well. In addition to identification of individual AE types that discriminate between treatment arms, it is also possible to assess overall safety of test product relative to placebo at interim looks. AdX as a measure of overall safety profile can be beneficial here because it is brief, cogent and inclusive. The question of interest is whether the picture of relative safety at interim look remains the same at the end of the trial. We shall use the data from completed trials to examine this question.

The total trial duration can be divided into (say) three equal parts and the summary measure of choice such as the count of AE types, AdX etc. can be calculated at each stage. This is illustrated with data from the GSK trial. Figure 1 shows AdX values at two interim and final time points, by gender and treatment.

AdX value appears to increase gradually as the trial progresses. This behaviour can be mainly attributed to the increase in count of AE types observed. At two-third of the trial period, we observe more than 80% of the AE types. Further, the gender-wise AdX values appear to reveal an interesting pattern as seen in the Figure 1.

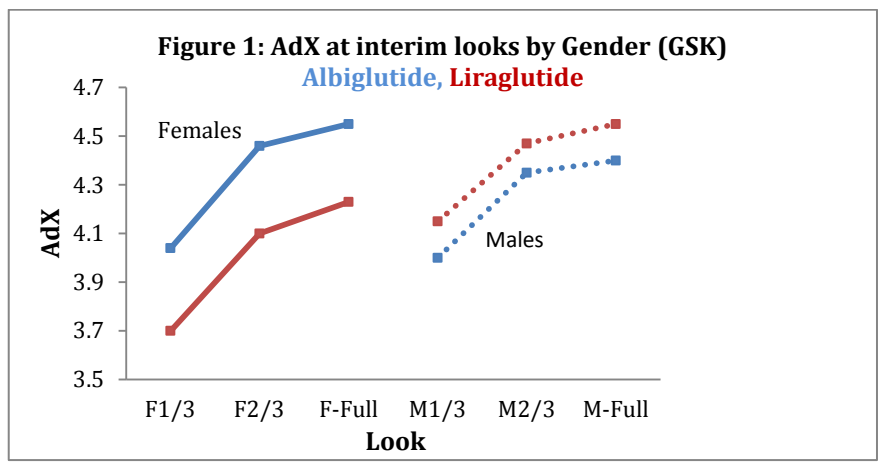

In the case of males (dotted lines), the AdX value for Liraglutide (comparator) is higher than that for Albiglutide (GSK drug) throughout; however, no difference is statistically significant. On the other hand, in the case of females (continuous lines) ordering of treatments reverses. Albiglutide has a higher AdX than Liraglutide at every look. Additionally, the difference is statistically significant every time. Thus, for females, the drug under investigation is less safe than the comparator. This fact would have been detected at the first interim look itself (one-third of the trial period).

These findings contradict the conclusion in CSR that *'Results from this study suggest that albiglutide administered once weekly has a comparable efficacy and safety profile to an approved GLP-1 agonist .....'* (CSR p.147)[14].

The above conclusion may be true for males but not for females. This shows how AdX value can be a useful guide to the DSMB in forming an opinion about the relative overall safety of different treatments at interim looks. It would also be relevant to check the extent to which such a guideline works for the other three trials.

In the EL trial the AE onset dates were not available in a large number of AE episodes. Hence the data could not be separated for interim looks. In the BI and RO trials, we observe a repetition of the suggested pattern i.e. the overall safety picture is reasonably clear at two-third of the trial.

## 4.7 Analysis by exposure to drug:

It appears that examination of safety by exposure can reveal interesting features of treatments. In oncology, exposure assessment is straightforward because the drug is administered sequentially in cycles and the number of cycles completed is a surrogate measure of exposure. The question of interest could be 'How does the AE profile in terms of AdX, number of AE types and number of AE episodes change with the number of chemotherapy cycles administered?'. In Figures 2A, 2B, and 2C number of cycles is represented on the X axis and the AE summary measure is represented on the Y axis for the RO trial.

We observe that the AdX values (Figure 2A) increase in the beginning but soon saturate. Here high dose has higher value of AdX throughout. The AdX curves for the other two arms overlap substantially. For the AE types (Figure 2B), although the picture is again of saturation, the placebo curve shows the lowest value throughout and the two active doses show higher values. The same holds for the number of episodes as well (Figure 2C). Saturation in AdX and the number of AE types is expected. However, saturation in the number of episodes appears to be counter-intuitive. The episode count is expected to increase with number of cycles received. The saturation is only in appearance; the cumulative count does not quite saturate but increases gently. The reason for gentle increase in count of episodes is that due to withdrawal or disease progression, the number of subjects declines sharply with increasing number of cycles. After about 15 cycles hardly any subjects are left (Figure 2D). The message from these graphs seems to be that picture of overall safety becomes clear after only about five or six cycles.

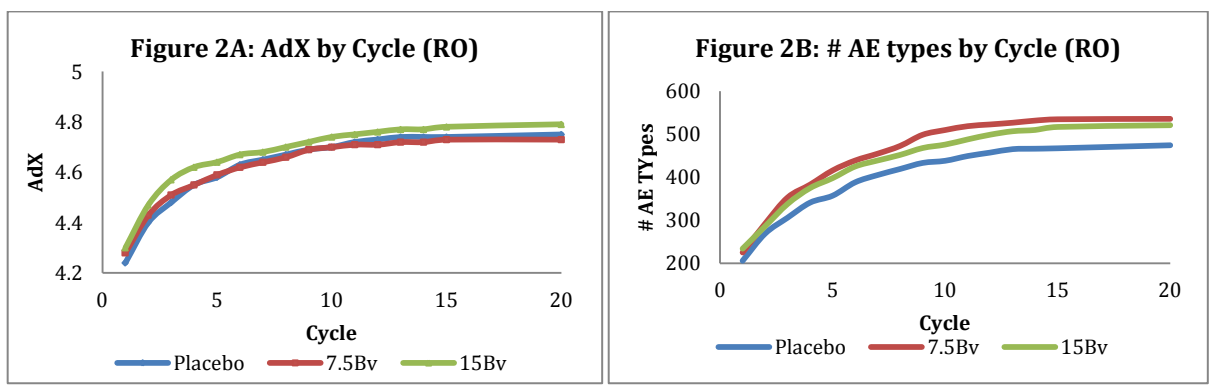

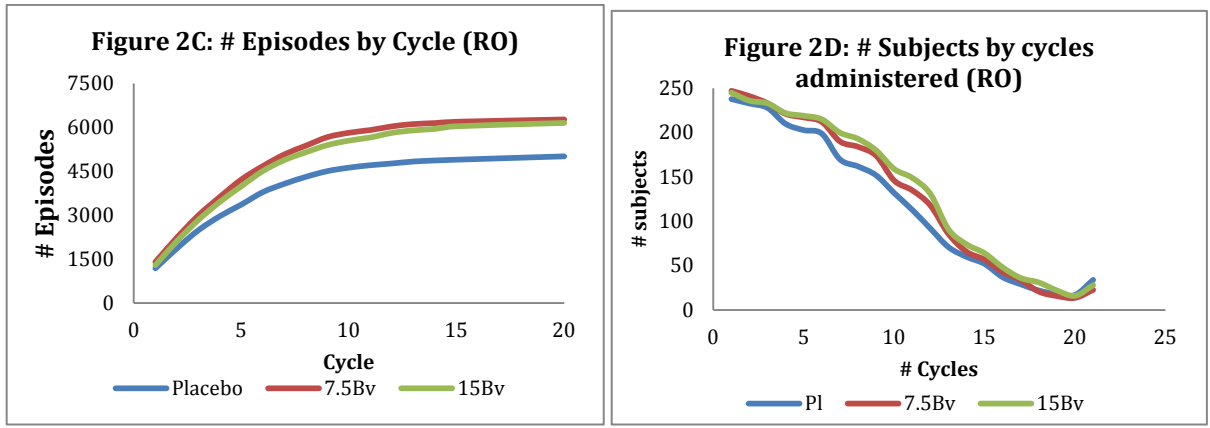

## 4.8 AdX with MedDRA Hierarchy:

So far AdX was assumed to be calculated at preferred term level (PT). This is not necessary. It can be calculated at any level in the MedDRA hierarchy. How will the index change with level of hierarchy chosen? We have the following general propositions. Firstly, as you go from preferred term (PT) to HLGT, we expect the AdX value to decline. The second proposition is that the rank of treatment in terms of AdX will remain the same across different levels. However, significance of differences in AdX values may change. So the third proposition is: if a difference in AdX values is significant at a higher MedDRA level, we expect it to remain significant at a lower level as well. Converse may not hold. It is possible to get significance at lower level while failing to get significance at higher level. Fourth and the last proposition is that if a treatment difference fails to attain significance at lower hierarchy level, it will continue to be so at higher level. Here also converse may not hold true. We illustrate these ideas with data on the BI trial. Figure 3 shows AdX values by treatment, background medication (Met or METSU) and MedDRA level (PT, HLT and HLGT).

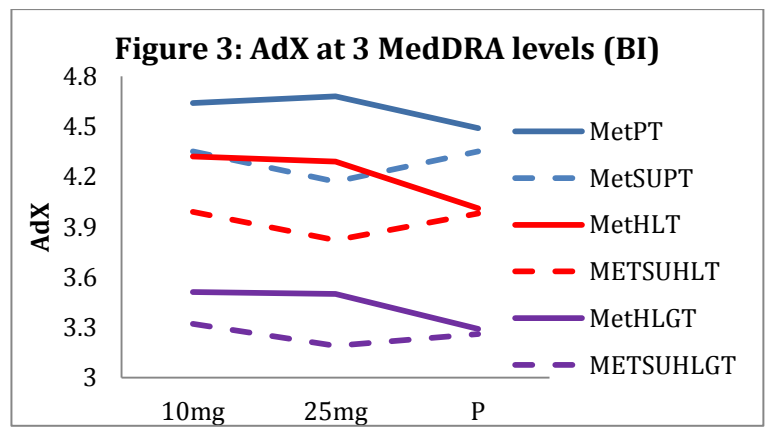

Each of six lines in Figure 3 represents AdX values for three treatments. Solid line is for background therapy 'Metformin' and dotted one for background therapy 'Metformin + SU'. As we go to higher level of MedDRA hierarchy, AdX value reduces (proposition one). This is because count of distinct types reduces. [Range for counts between different treatment groups: PT level (150,200), HLT level (120,130), HLGT level (70, 80)]. Further, three lines for each background therapy are roughly parallel implying ranking remains unchanged irrespective of hierarchy level (proposition two). Table 11 is useful as illustration of propositions three and four. It gives p-values for differences (Active- placebo) for two background therapy groups.

| Table 11:Treatment comparison by MedDRA levels (p-values) | | | | |
|---|---|---|---|---|
| | Metformin | | Met + SU | |
| 1 | 2 | 3 | 4 | 5 |
| MedDRA level | 10mg-Placebo | 25mg-Placebo | 10mg-Placebo | 25mg-Placebo |
| PT | 0.042 | 0.013 | 0.500 | 0.039 |
| HLT | 0.000 | 0.002 | 0.457 | 0.050 |
| HLGT | 0.016 | 0.022 | 0.258 | 0.232 |
| Data Source: NCT01159600; Sponsor Name: Boehringer Ingelheim (BI); Indication: Diabetes Mellitus Type II. | | | | |

Consider column 2. Here p-value for HLGT is small. Proposition three predicts that difference should be significant at HLT and PT as well. This is true. Same result holds for column 3. Consider column 4. P-value for PT is large. Proposition four predicts that p-values for HLT and HLGT should also be large. This is true. Column 5 demonstrates that converse of proposition three is not necessarily true. The difference is significant at PT and HLT level but not at HLGT level.

### 4.9 Using AdX in benefit risk analysis:

Drug approval typically requires a confirmatory proof of efficacy followed by a benefit-risk assessment. The latter is another hot topic in the clinical trial domain. It is felt that *'regulators need to refine their methods of assessing benefit–risk balances and switch from 'implicit' to 'explicit' decision making'* [15]. Benefit-risk ratio (BRR) is discussed extensively in literature[16]. Several measures of BRR have been proposed[17,18]. But there is no unanimity. The key problem appears to be the quantification of risk. Broad summary terms such as 'generally safe' or 'well tolerated' are commonly used. These terms have been criticized in literature[3]. Phrases such as 'favourable risk- benefit profile', 'manageable toxicity' may be adequate to describe the safety profile verbally but not so for calculating the BRR. Analysis often begins with the application of statistical methods to an individual efficacy end point or an adverse event type. However, overall conclusion is only a judgmental statement. This is because of difficulty in aggregating component results.

Our approach is to avoid these pitfalls. We can use a primary efficacy end point or a suitable composite of multiple efficacy end points to represent the benefit. AdX provides a measure of risk. The **R**atio of **E**fficacy to **Ad**versity (REAd) can be used as the BRR of a treatment (REAd= average efficacy/AdX). For a comparison of two treatments, the ratio of the two REAds can be calculated. This ratio can be called the **Re**lative REAd (Re-REAd). If two treatments T1 and T2 have similar efficacy, but T1 has a smaller AdX (safer) than T2, then the REAd value of T1 will be higher. We illustrate these ideas by applying them to the EL trial.

Here is the conclusion from the sponsors of the EL trial: *'Patients on GT arm had a statistically significant improvement in TtDPD with an approximate 50% relative increased probability of a patient being documented progression free at 6 months ... This was accompanied with a statistically significant improvement in PFS and overall tumour response rate for the patients on GT arm. ... In addition, the patients on the GT arm had a significant improvement in overall valuation of life at cycles 5 and 6 compared with baseline. Overall, GT combination chemotherapy is an effective and well tolerated chemotherapy regimen, with an expected and manageable toxicity. This favourable benefit risk profile supports the use of Gemcitabine + Paclitaxel combination therapy in patients with metastatic breast cancer'* (CSR p. 6).

Treatments cannot be compared only in terms of efficacy unless proper discounting for safety performance is applied [1]. The sponsors have argued that *'No new trends or safety concerns were observed following examination of the adverse events reported during this study'*. In this sense, safety profiles of two arms are deemed to be similar. Hence, efficacy of GT relative to T (control) essentially represents the benefit risk balance. The conclusion that safety profiles are similar appears to be based on the rates of occurrence of the events of special interest. It does not seem to consider the entire gamut of adverse events. If the overall safety profiles of the two arms are different, then we must take cognizance of differences in safety and discount benefit accordingly.

Table 12 shows the efficacy and REAd values by treatment. The conclusion based on Re-REAd is in agreement with that of the sponsors, viz. the performance of GT is slightly better than that of T. Thus, in this case, the proposed BRR measure confirms the conclusions in the CSR. Of course statistical testing of the hypothesis that two treatments have comparable BRR (Re-REAd=1) has to be addressed. One possible solution is using boot strap confidence interval for Re-REAd.

| Table 12: Calculation of REAd and Re-REAd(EL) | | | |
|---|---|---|---|
| | Efficacy | Safety | |
| | Median PFS (months) | AdX | REAd* |
| GT | 5.3 | 3.64 | 1.46 |
| T | 3.5 | 3.48 | 1.01 |
| Re-REAd(GT/T)** | | | 1.45 |
| *REAd: **R**atio of **E**fficacy and **AdX** , ** Re-REAd: Relative REAd | | | |
| Data Source: NCT00006459; Sponsor Name: Eli Lilly; Indication: Breast Cancer. † Treatment**:** GT= Gemcitabine + Paclitaxel; T= Paclitaxel. | | | |

The next illustration reveals some finer points of treatment comparison for the BI trial, unnoticed in the CSR. The results of this analysis are shown in Table 13.

| Table 13: Benefit Risk Analysis for Two Sub-studies in BI Trial | | | | | |
|---|---|---|---|---|---|
| Sub Study | Row number | Statistic | Treatment Arm | | |
| | | | 10 mg | 25mg | Placebo |
| Metformin | 1 | Benefit: Primary Efficacy: MeanCFB HbA1C | -0.72 | -0.75 | -0.13 |
| | 2 | Risk: AdX | 4.64 | 4.68 | 4.49 |
| | 3 | REAd | 0.155 | 0.160 | 0.029 |
| | 4 | Re-REAd (Empagliflozin/Placebo) | 5.36 | 5.54 | |
| Metformin + SU | 5 | Benefit:Primary Efficacy: MeanCFB HbA1C | -0.80 | -0.77 | -0.18 |
| | 6 | Risk: AdX | 4.35 | 4.17 | 4.35 |
| | 7 | REAd | 0.184 | 0.185 | 0.041 |
| | 8 | Re-REAd(Empagliflozin/Placebo) | 4.44 | 4.46 | |
| Data Source: NCT01159600; Sponsor Name: Boehringer Ingelheim (BI); Indication: Diabetes Mellitus Type II. | | | | | |

For each sub-study and treatment arm, Table 13 shows the efficacy (borrowed from the CSR), risk (AdX), benefit-risk ratio (REAd) and Re-REAd of the active treatment Vs the placebo. In the metformine sub-study, we observe that both the treatment arms show substantially larger efficacy than the placebo. On the other hand the differences in risk are much smaller. Hence our index for benefit-risk balance viz. REAd is quite high for both active treatment arms. The story with Metformin + SU is essentially the same.

In conclusion we can say that irrespective of the background medication, each active treatment arm achieves better performance than the placebo (background medication alone).

The last point is comparison of sub-studies. Here we are going beyond the report of the trial. Naturally CSR is silent about this. We attempt to address the following question: How does the performance of an active treatment compare across background medication? The efficacy value for each treatment arm is slightly higher in the Metformine+SU sub-study than corresponding value in the Metformine sub-study. The AdX value for each treatment arm is slightly lower in the Metformine+SU sub-study than corresponding value in the Metformine sub-study. Therefore the Metformine +SU sub-study shows higher REAd values for each treatment arm. It is tempting to conclude that active treatment performs better with Metformine + SU background therapy. The unexpected twist comes next.

The above conclusion may be challenged using the argument that the placebos in two sub-studies are not identical. Perhaps we should rephrase our main question. Is the improvement shown by active treatment over the placebo similar for two background medications? The ratio Re-REAd (active /placebo) helps in answering this question. These Re-REAd values are shown in rows 4 and 8 of Table 13. The Re-REAd values are higher when Metformine is the background medication compared with corresponding values, when Metformine+SU as the background medication. Thus, each dose of active treatment provides a better advantage over the placebo when Metformine is the back ground therapy than when Metformine+ SU is the background therapy. Such analysis became possible because of quantification of risk (AdX).

## 5. Discussion

This paper proposes a new index to summarise safety profile of a treatment in a clinical trial. This index is not intended to replace present method of focussing on events of special interest which are clinically important. If a treatment is judged to be unsafe and unacceptable based on rates of occurrence of these events, other adverse events may not play a role in decision making. Otherwise AdX is thought to be a more effective tool than present method of reporting just counts and percentages. The paper demonstrates how hypothesis of comparable safety profile can be statistically tested. It further shows use of this index in subgroup analysis and benefit risk balance assessment. Such analysis can lead to findings/ features likely to be missed in conventional analysis.

**Acknowledgement:**

This work would not have been possible without access to some completed clinical trials. We are grateful to Eli Lilly, Boehringer Ingelheim, Glaxo-Smith-Klein and Hoffman La Roche for making trial data available through https://www.clinicalstudydatarequest.com

We are thankful to Ranganath Nayak, Cyrus Mehta, Nitin Patel, Ajay Sathe, Jim Bolognese, Yannis Jemiai, Steve Herbert, Chuck Gelbs, Chris Schoonmaker and Joe Heyse for useful discussions and encouragement.

The hard work of processing the data and generating tables and figurers was done by our technical team consisting of Adarsh Nagare, Sanhita Yeolekar, Ritika Yadav, Vaishnavi Bhambure and Sanjanarao Lade.

**References:**

1. Woodcock, J. *CDER DRUG SAFETY PRIORITIES 2015-2016* (p. 5) https://www.fda.gov/downloads/Drugs/DrugSafety/UCM523486.pdf (date accessed March 9, 2018)
2. Schactman, M. and Wittes, J. Why a DMC Safety Report Differs from a Safety Section Written at the End of the Trial, In: Qi J. and Xia, H., eds Quantitative Evaluation of Safety in Drug Development: Design, Analysis and Reporting. Boca Raton, FL: CRC Press 2015:69-92
3. Lineberry, N., Berlin, J., Mansi, B. et al. Recommendations to improve adverse event reporting in clinical trial publications: a joint pharmaceutical industry/journal editor perspective *BMJ 2016:*355, doi: https://doi.org/10.1136/bmj.i5078
4. Mehrotra, D. and Adewale, A. Flagging clinical adverse experiences: reducing false discoveries without materially compromising power for detecting true signals, *Statist. Med*. 2012: 31: 1918–1930.
5. Yeh, S. Clinical Adverse Events Data Analysis and Visualization (2007) http://www.lexjansen.com/pharmasug/2007/po/PO10.pdf (accessed on 9th March, 2018)
6. Ohad A., Heiberger R. and Lane P. Graphical Approaches to the Analysis of Safety Data from Clinical Trials, *Pharmaceut. Statist*. 2008; 7: 20–35 DOI: 10.1002/pst.254
7. Qi J. and Xia, H., eds Quantitative Evaluation of Safety in Drug Development: Design, Analysis and Reporting. Boca Raton, FL: CRC Press 2015:ix
8. Ke, C., Qi J., and Snapinn, S. Risk-Benefit Assessment Approaches, In: Qi J. and Amy Xia, eds. Quantitative Evaluation of Safety in Drug Development Design, Analysis and Reporting, Boca Raton, FL: CRC Press 2015:267-288.

9. Pocock S., The pros and cons of noninferiority trials, *Fundamental & Clinical Pharmacology:* 2003: 17: 483–490.

10. Packer, M. (2008) https://www.fda.gov/ohrms/dockets/ac/08/slides/2008-4387s1-04-Packer.pdf (accessed on 9th March, 2018).

11. Khaled E. and Arbuckle L. Anonymizing Health Data: Case Studies and Methods to Get You Started. Sebastopol, CA: O'Reilly Media, Inc 2014:33.

12. Paninski L. Estimation of Entropy and Mutual Information. *Neural Computation* 2003:15: 1191–1253.

13. Taylor, G. and Wainwright, P. Open label extension studies: research or marketing? *BMJ* 2005: 331(7516): 572–574. doi: 10.1136/bmj.331.7516.572.

14. *https://www.gsk-clinicalstudyregister.com/study/114179?search=study&study_ids=114179#csr* (checked on March 14, 2018)

15. Eichler, H., Abadie, E., Raine, J., et al . Safe Drugs and the Cost of Good Intentions *N. Engl. J. Med;* 2009: 360:1378-1380 doi: 10.1056/NEJMp0900092

16. Loke Y., Price D., and Herxheimer A. Systematic reviews of adverse effects: framework for a structured approach *BMC Medical Research Methodology* 2007:32 https://doi.org/10.1186/1471-2288-7-32.

17. Qi, J., and He W., eds. Benefit-Risk Assessment Methods in Medical Product Development Bridging Qualitative and Quantitative Assessments, Chapman and Hall.2016: 86.

18. Sashegyi,A., Felli, J., and Noel, R., eds. Benefit-Risk Assessment in Pharmaceutical Research and Development, Boca Raton, FL. Chapman and Hall. 2014:5.